\documentclass[12pt]{article}
\usepackage{amssymb,graphicx}
\usepackage{epstopdf}
\usepackage{amsmath,amsfonts}
\usepackage{epsfig}
\usepackage[dvipsnames]{xcolor}

\newcommand{\be}{\begin{equation}}
\newcommand{\ee}{\end{equation}}
\newcommand{\bea}{\begin{eqnarray}}
\newcommand{\eea}{\end{eqnarray}}
\newcommand{\bg}{\begin{gather}}
\newcommand{\eg}{\end{gather}}
\newcommand{\bseq}{\begin{subequations}}
\newcommand{\eseq}{\end{subequations}}

\usepackage{hyperref}
\definecolor{linkcolor}{HTML}{799B03}
\definecolor{urlcolor}{HTML}{799B03}
\hypersetup{pdfstartview=FitH,linkcolor=linkcolor,urlcolor=urlcolor,colorlinks=true}

\begin{document}

%\vspace{10pt}

\begin{center}
{\LARGE \bf  Genesis with general relativity asymptotics in beyond Horndeski theory}

\vspace{20pt}
S. Mironov$^{a,c,d}$\footnote{sa.mironov\_1@physics.msu.ru},
V. Rubakov$^{a,b}$\footnote{rubakov@inr.ac.ru},
V. Volkova$^{a}$\footnote{volkova.viktoriya@physics.msu.ru}
\renewcommand*{\thefootnote}{\arabic{footnote}}
\vspace{15pt}

$^a$\textit{Institute for Nuclear Research of the Russian Academy of Sciences,\\
60th October Anniversary Prospect, 7a, 117312 Moscow, Russia}\\
\vspace{5pt}

$^b$\textit{Department of Particle Physics and Cosmology, Physics Faculty,\\
M.V. Lomonosov Moscow State University,\\
Vorobjevy Gory, 119991 Moscow, Russia}

$^c$\textit{Institute for Theoretical and Experimental Physics,\\
Bolshaya Cheriomyshkinskaya, 25, 117218 Moscow, Russia}

$^d$\textit{Moscow Institute of Physics and Technology,\\
Institutski pereulok, 9, 141701, Dolgoprudny, Russia}
\end{center}

\vspace{5pt}

\begin{abstract}
  We suggest a novel version of a cosmological Genesis model
  within beyond Horndeski theory. It combines the initial Genesis behavior
  of Creminelli et al.~\cite{Creminelli1,Creminelli2} with the
  complete stability property of the previous beyond Horndeski
  construction~\cite{bounce}.
  The specific features of the model are that
space-time rapidly tends to Minkowski in the asymptotic past and
  that both the asymptotic
past and future are described by General Relativity (GR).
\end{abstract}

%%%%%%%%%%%%%%%%%%%%%%%%%%%%%%%%%%%%%%%%%%%%%%%%%%%%%%%%%%%%%%%%%%%%%%%%%
\section{Introduction}

The model of the Universe starting with the Genesis epoch
of
nearly flat space-time and growing energy density and expansion rate,
is an
example of non-standard cosmology
based on the violation of the Null Energy Condition (NEC)
(for a review see,  e.g., Ref.~\cite{RubakovNEC}) or, more generally,
the Null Convergence Condition (NCC)~\cite{NCC}.
The Genesis scenario~\cite{Creminelli1} was
first suggested within a simple class of conformal
Galileon theories minimally coupled to gravity,
where growing energy density ($\dot{\rho} > 0$) does not
necessarily lead to  instabilities.
In fact, it was later shown that there is a much wider class of
scalar-tensor theories with a similar mechanism of safe NEC/NCC violation
-- generalized Galileon theories or, equivalently,
Horndeski theories~\cite{Horndeski,Deffayet}.

Horndeski theories are  general
scalar-tensor gravities with second order
equations of motion. These have been
further generalised to theories with higher order equations of motion,
dubbed DHOST
theories~\cite{Zuma,Gleyzes,Langlois1,Ben,Langlois2,LangloisRev}.
The constraint structure
of the DHOST theories
is such that they propagate
only three dynamical degrees of freedom, just like Horndeski
theories. Horndeski theories and their generalizations are
an interesting playground for studying stable 
NEC/NCC-violating cosmologies
(for a review see,  e.g., Ref.~\cite{KobayashiRev}), and
Genesis in particular~\cite{gen1,gen2,gen3}.

One of the main reasons for going beyond Horndeski, at least in the
context of early cosmology, is to construct examples of complete
spatially flat,
non-singular cosmological scenarios like Genesis. Modulo options that are
dangerous from the viewpoint of geodesic completeness and/or
strong coupling~\cite{LMR,Koba,Ijjas:2016tpn} (see, however~\cite{VOY}), Horndeski theories are not
suitable for this purpose
because of the inevitable development of gradient or ghost instabilities
at some stage of the
evolution~\cite{LMR,Koba,Kolevatov:2016ppi,Akama:2017jsa}.
However, this no-go theorem does not apply to DHOST theories,
as demonstrated in Refs.~\cite{CaiEFT,CreminelliEFT,bounce}
for a subclass usually
referred to as "beyond Horndeski" (aka GLVP~\cite{Gleyzes}).
Indeed, this subclass has been  used for constructing
non-singular cosmological models of the bouncing Universe  and
Genesis, which are stable at the linearised level during the entire
evolution~\cite{bounce,CaiBounce,bounceI}.

Previous constructions of complete bouncing and Genesis models
in beyond Horndeski theories were limited by overestimating
the danger of a phenomenon called $\gamma$-crossing (or $\Theta$-crossing).
The discussion of this phenomenon is fairly technical, and we
postpone it to Section~\ref{sec:stable}. It suffices to point
out here that insisting on the absence of $\gamma$-crossing
prevents one from constructing
bounce and Genesis models
where linearized gravity agrees
with GR both in the asymptotic future and in the asymptotic past, and, in
the Genesis case, whose space-time rapidly tends
to Minkowski in the asymptotic past. An example
is a Genesis-like model of Ref.~\cite{bounce} where the scale factor behaves as
$a(t) \propto |t|^{-1/3}$ as $t \to -\infty$.

It has been shown, however, that $\gamma$-crossing is, in fact,
an innocent phenomenon. Originally, this fact
was established in Newtonian gauge~\cite{Ijjas:2017pei} and then
confirmed in unitary gauge~\cite{bounceI}. It opens up the possibility
to construct new bouncing and Genesis models\footnote{We point out,
  however, that the no-go theorem is valid in Horndeski theories
  irrespectively of $\gamma$-crossing.}. Indeed, an example of
a fully stable, spatially flat
bouncing model has been constructed in beyond Horndeski
theory~\cite{bounceI}, whose asymptotic past and future are described,
modulo small corrections, by GR with a conventional massless scalar
field.

In this paper we continue along this line and
suggest an example of a complete, stable cosmological Genesis model
in a theory of beyond Horndeski subclass.
In our model, the Universe starts from the asymptotic
Minkowski state and undergoes the Genesis stage
at early times, which is very similar to the subluminal
version of the original Genesis scenario
in Ref.~\cite{Creminelli2}.
The specific feature of the model is that the driving field starts off
as cubic Galileon (and hence gravity is described by
GR modulo small corrections), turns,
as the system evolves, into beyond Horndeski type and
 becomes, in the asymptotic future, a canonical massless scalar field
in GR.
The model is constructed so that there are neither ghosts
nor gradient instabilities about the background at all times,
i.e. the solution is completely stable. We also ensure that
the propagation of both
scalar and tensor perturbations  is
subluminal (or luminal at most) during entire evolution.
All these features are obtained by a judicial choice
of
the beyond
Horndeski Lagrangian. Our example thus shows that beyond Horndeski
theories are capable of yielding Genesis models with
fairly simple properties, which
may be advantageous for constructing
realistic early Universe models.

The paper is organized as follows. We briefly revisit
basic formulas
of the linearized perturbation theory for
(beyond) Horndeski theories in Sec.~\ref{sec:stable}.
There, we also discuss the
$\gamma$-crossing
phenomenon and its role in the no-go theorem.
In Sec.~\ref{sec:genesis} we reconstruct the beyond Horndeski
Lagrangian which admits a completely healthy Genesis solution with
GR asymptotics and explicitly demonstrate that the solution is stable.
We conclude in Sec.~\ref{sec:conclusion}.

%%%%%%%%%%%%%%%%%%%%%%%%%%%%%%%%%%%%%%%%%%%%%%%%%%%%%%%%%%%%%%%%%%%%%%%%%
\section{Stability of  cosmological backgrounds in beyond Horndeski theory}
\label{sec:stable}

In this section we introduce the notations and revisit several known
results related to the stability analysis of homogeneous
cosmological solutions in beyond Horndeski theory.

We consider the quartic subclass of beyond Horndeski
theory with the following action
(mostly negative signature):
\begin{multline}
\label{lagrangian}
S = \int d^4x \sqrt{-g}\Big(F(\pi,X) + K(\pi,X)\Box\pi - G_4(\pi,X)R \\
+ \left(2 G_{4X}(\pi,X) + F_4 (\pi,X) \; X\right)\left[\left(\Box\pi\right)^2-\pi_{;\mu\nu}\pi^{;\mu\nu}\right] \\
- 2 F_4 (\pi,X) \left[\pi^{,\mu} \pi_{;\mu\nu} \pi^{,\nu}\Box\pi -  \pi^{,\mu} \pi_{;\mu\lambda} \pi^{;\nu\lambda}\pi_{,\nu} \right]\Big),
\end{multline}
where $\pi$ is the Galileon scalar
field,
$X= g^{\mu\nu}\pi_{,\mu}\pi_{,\nu}$,
$\pi_{,\mu}=\partial_\mu\pi$,
$\pi_{;\mu\nu}=\triangledown_\nu\triangledown_\mu\pi$,
$\Box\pi = \triangledown_\nu\triangledown^\nu\pi$,
$G_{4X}=\partial G_4/\partial X$.
Let us emphasize that the function $F_4(\pi,X)$ is characteristic of beyond
Horndeski theory, whereas $F_4=0$ in Horndeski subclasses.
The corresponding Einstein equations for a
flat FLRW background read
\begin{subequations}
\label{einstein}
\begin{align}
\delta g^{00}:\quad
&F-2F_XX-6HK_XX\dot{\pi}+K_{\pi}X+6H^2G_4+6HG_{4\pi}\dot{\pi}
\\\nonumber&-24H^2X(G_{4X}+G_{4XX}X)+12HG_{4\pi X}X\dot{\pi}
%\\\nonumber&
-6H^2X^2(5F_4+2F_{4X}X)=0,\\
%%%%%%%%
\delta g^{ii}:\quad
&F-X(2K_X\ddot{\pi}+K_\pi)+2(3H^2+2\dot{H})G_4-12H^2G_{4X}X-8\dot{H}G_{4X}X
\\\nonumber&-8HG_{4X}\ddot{\pi}\dot{\pi}-16HG_{4XX}X\ddot{\pi}\dot{\pi}+2(\ddot{\pi}+2H\dot{\pi})G_{4\pi}+2XG_{4\pi\pi}
\\\nonumber&+4XG_{4\pi X}(\ddot{\pi}-2H\dot{\pi})-2F_4X(3H^2X+2\dot{H}X+8H\ddot{\pi}\dot{\pi})
\\\nonumber&-8HF_{4X}X^2\ddot{\pi}\dot{\pi}
%\\\nonumber&
-4HF_{4\pi}X^2\dot{\pi}=0.
\end{align}
\end{subequations}

In what follows, we carry out a
stability analysis about flat FLRW
background
and adopt the standard  parametrization of
perturbations:
\be
\label{perturbations}
ds^2 = (1+2\alpha)dt^2 - \partial_i\beta \;dt dx^i - a^2 (1 + 2\zeta \delta_{ij} + 2 \partial_i\partial_j E + h_{ij}^T ) dx^i dx^j,
\ee
where $\alpha$, $\beta$, $\zeta$ and $E$ belong to a scalar sector,
while $h_{ij}^T$ denotes transverse traceless tensor perturbations.
We adopt the unitary gauge approach, where both the longitudinal perturbation $E$ and the scalar field perturbation vanish, $E = \delta\pi = 0$.

The unconstrained form of the quadratic action
in terms of tensor modes $h_{ij}^T$ and curvature perturbation $\zeta$
reads (see, e.g., Refs.~\cite{Kobayashi,bounce,KobayashiRev} for
a detailed derivation):
\be
\label{quadr_action}
S^{(2)}=\int\mathrm{d}t\mathrm{d}^3x a^3\left[\dfrac{\mathcal{{G}_T}}{8}\left(\dot{h}^T_{ij}\right)^2-\dfrac{\mathcal{F_T}}{8a^2}\left(\partial_k h_{ij}^T\right)^2+\mathcal{G_S}\dot{\zeta}^2-\mathcal{F_S}\dfrac{(\triangledown\zeta)^2}{a^2}\right],\\
\ee
where the coefficients involved are
\begin{subequations}
\label{gsfs}
\begin{align}
\label{eq:GS_setup}
&\mathcal{G_S}=\dfrac{\Sigma\mathcal{{G}_T}^2}{\Theta^2}+3\mathcal{{G}_T},\\
\label{eq:FS_setup}
&\mathcal{F_S}=\dfrac{1}{a}\dfrac{\mathrm{d}\xi}{\mathrm{d}t}-\mathcal{F_T},\\
\label{eq:xi_func_setup}
&\xi=\dfrac{a\left(\mathcal{{G}_T}+\mathcal{D}\dot{\pi}\right)\mathcal{{G}_T}}{\Theta},
\end{align}
\end{subequations}
and
\begin{subequations}
\label{list_coeff_setup}
\begin{align}
\label{eq:Gt_coeff_setup}
&\mathcal{G_T}=2G_4-4G_{4X}X - 2F_4X\dot{\pi},\\
%%%%
\label{eq:Ft_coeff_setup}
&\mathcal{F_T}=2G_4,\\
%%%%
\label{eq:D_coeff_setup}
&\mathcal{D}=2F_4X\dot{\pi},\\
%%%%
%%%%
\label{eq:Theta_coeff_setup}
&\Theta=-K_XX\dot{\pi}+2G_4H-8HG_{4X}X-8HG_{4XX}X^2+G_{4\pi}\dot{\pi}+2G_{4\pi X}X\dot{\pi}\\
\nonumber
&-10HF_4X^2-4HF_{4X}X^3,\\
%%%%
\label{eq:Sigma_coeff_setup}
&\Sigma=F_XX+2F_{XX}X^2+12HK_XX\dot{\pi}+6HK_{XX}X^2\dot{\pi}-K_{\pi}X-K_{\pi X}X^2\\
\nonumber&-6H^2G_4+42H^2G_{4X}X+96H^2G_{4XX}X^2+24H^2G_{4XXX}X^3-6HG_{4\pi}\dot{\pi}\\
% \nonumber&\\
\nonumber&-30HG_{4\pi X}X\dot{\pi}-12HG_{4\pi XX}X^2\dot{\pi}+90H^2F_4X^2+78H^2F_{4X}X^3\\
\nonumber&+12H^2F_{4XX}X^4.
\end{align}
\end{subequations}
The explicit form of coefficients~\eqref{list_coeff_setup}
is given for the Lagrangian in~\eqref{lagrangian}.
The issue of gradient instabilities is governed by coefficients
$\mathcal{F_T}$ and $\mathcal{F_S}$, while the signs of
$\mathcal{G_T}$ and $\mathcal{G_S}$ indicate whether there are ghosts
in the linearized theory. A fully stable background is such that
$\mathcal{F_T}, \mathcal{F_S}, \mathcal{G_T}, \mathcal{G_S} > 0$.
The propagation speeds squared for tensor and scalar modes in
the quadratic action~\eqref{quadr_action} are, respectively,
\begin{equation}
\label{speed}
c_\mathcal{T}^2=\dfrac{\mathcal{F_T}}{\mathcal{{G}_T}},\qquad c_\mathcal{S}^2=\dfrac{\mathcal{F_S}}{\mathcal{G_S}}.
\end{equation}
By requiring that the propagation is not superluminal,
we write the stability conditions as follows:
\be
\label{stability_constraints}
\mathcal{{G}_T} \geq \mathcal{{F}_T}> \epsilon > 0, \qquad \mathcal{{G}_S} \geq \mathcal{{F}_S}> \epsilon  > 0.
\ee
Introduction of a positive constant
$\epsilon$ in the conditions~\eqref{stability_constraints} is meant to
avoid
a potential strong coupling issue
(see Refs.~\cite{bounceI,bounceQ,Mironov} for discussion).

One point to keep in mind when constructing cosmological models is
the form of the
stability condition $\mathcal{{F}_S} > 0$,
which constrains the behaviour of $\xi$
(see eqs.~\eqref{eq:FS_setup} and~\eqref{eq:xi_func_setup})
\be
\label{nogo1}
\dfrac{\mathrm{d}\xi}{\mathrm{d}t}   > \epsilon + \mathcal{{F}_T}> 0 \; ,
\;\;\;\;\; \xi =\dfrac{a\left(\mathcal{{G}_T}+\mathcal{D}\dot{\pi}\right)\mathcal{{G}_T}}{\Theta}  \; .
%\dfrac{\mathrm{d}\xi}{\mathrm{d}t} = \dfrac{\mathrm{d}}{\mathrm{d}t}\left(\dfrac{a\left(\mathcal{{G}_T}+\mathcal{D}\dot{\pi}\right)\mathcal{{G}_T}}{\Theta}\right) > \epsilon + \mathcal{{F}_T}> 0 \; .
\ee
It reveals
 the crucial role of the beyond Horndeski
 coefficient $\mathcal{D}$: for  $\mathcal{D}=0$ (Horndeski case),
 growth of $a\mathcal{{G}_T}^2 /\Theta$ forbids a complete, stable
 bouncing Universe
 and Genesis, which is precisely the no-go theorem~\cite{Koba}.

Another subtle issue has to do with the function $\Theta$
in~\eqref{nogo1}.
As  shown in Refs.~\cite{bounce,bounceI,bounceQ,Mironov}
the adjustment of $\Theta$ does not help with evading
the no-go theorem, yet $\Theta$ becomes important when it comes
to  asymptotics as $t \to \pm\infty$.
Namely, if one insists, as we do in this paper, that
space-time is asymptotically flat in the asymptotic past,
and
linearized gravity reduces to
GR in both asymptotics, then $\Theta$ must cross zero sometime in between.
The reason for this is that these asymptotics are obtained with
$F_4 \to 0$ as $t \to \pm \infty$, which in turn gives
$\mathcal{D} \to 0$  as $t \to \pm \infty$. Now, since $\dot{\xi} > \epsilon
> 0$ at all times, we have $\xi < 0$ as $t\to -\infty$ and
$\xi > 0$ as $t\to +\infty$. With $\mathcal{D} \to 0$  as $t \to \pm \infty$,
this means that  $\Theta < 0$ as $t\to -\infty$ and
$\Theta > 0$ as $t\to +\infty$ (this is confirmed by an explicit example
below), implying that $\Theta$ crosses zero at some
finite $t$. Note that the function $\Theta$ is denoted by $\gamma$
in Ref.~\cite{Ijjas:2016tpn}, so the phenomenon we are talking about
is called $\gamma$-crossing.

At a glance, eqs.~\eqref{gsfs} suggest that both $\mathcal{G_S}$ and
$\mathcal{F_S}$ blow up as $\Theta$ crosses zero. That was the reason,
for instance, for requiring that  $\Theta$ does not cross zero in bouncing
and Genesis-like models in Ref.~\cite{bounce}. In full accordance with
the above argument, non-vanishing $\Theta$ resulted in
non-trivial asymptotic theory of beyond Horndeski type at early times,
which was grossly different from GR
(see also Ref.~\cite{bounceI} for further discussion).

However, the analytical forms of $\mathcal{G_S}$ and $\mathcal{F_S}$
in eqs.~\eqref{gsfs} suggest that the dispersion relation
$c_\mathcal{S}^2 = \mathcal{F_S}/\mathcal{G_S}$ is finite at
$\gamma$-crossing, which implies that the scalar sector remains healthy.
Indeed, it was shown by Ijjas~\cite{Ijjas:2017pei} that
equations for perturbations are non-singular in Newtonian gauge.
Furthermore, it was explicitly checked in Ref.~\cite{bounceI}
that $\gamma$-crossing
does not lead to singularities of
solutions for $\zeta$,
and hence does not cause any trouble in stability analysis.
A completely healthy bouncing model with both asymptotics
described by a massless scalar field + GR was suggested
in Ref.~\cite{bounceI}, where it was shown that
$\gamma$-crossing is crucial for the model to be consistent.

In the next Section we also allow for $\gamma$-crossing
and construct a Genesis model whose initial stage coincides with
the original subluminal Genesis~\cite{Creminelli2}, while
the asymptotic future is described by GR with a
canonical
massless scalar field. In between these
stages the theory is essentially of beyond Horndeski type, which
ensures that the no-go theorem for non-singular
cosmologies is circumvented.

%%%%%%%%%%%%%%%%%%%%%%%%%%%%%%%%%%%%%%%%%%%%%%%%%%%%%%%%%%%%%%%%%%%%%%%%%
\section{Stable subluminal Genesis: an example}
\label{sec:genesis}

We make use of the reconstruction procedure, which has
proven  efficient
in constructing
other types of completely
stable non-singular cosmological solutions in beyond Horndeski
theories~\cite{bounce,bounceI}. Namely, we choose a specific form
of the Hubble parameter $H(t)$ and Galileon field $\pi(t)$
and reconstruct
the Lagrangian functions
by making use of the stability conditions and background field equations,
along with the additional constraints on the
asymptotic behaviour of the theory
as $t \to \pm \infty$.

For the sake of simplicity we consider a monotonously growing scalar
field $\pi$ with the following time dependence:
\be
\label{rolling_pi}
\pi (t) = t, \quad X=1,
\ee
which can always be obtained by field redefinition.

In our example, we assume that the initial Genesis stage is the
same as in the
subluminal version~\cite{Creminelli2}
of the original Genesis~\cite{Creminelli1}.
Hence, the early time asymptotic of $H(t)$ is
\be
\label{Hubble_early}
t \to -\infty: \quad
H = \dfrac{f^3}{4\Lambda^3} \dfrac{\left(1+\frac{\alpha}{3}\right)}{(-t)^3} , \quad
a(t) = 1 + \dfrac{f^3}{8\Lambda^3} \dfrac{\left(1+\frac{\alpha}{3}\right)}{(-t)^2},
\ee
and  the
Lagrangian is
\be
\label{lagr_asymp_early}
\mathcal{L}_{t\to-\infty} = -\dfrac12 R -\dfrac{3 f^3}{4 \Lambda^3} \dfrac{(1+\alpha)}{\pi^4} \cdot X
+ \dfrac{3 f^3}{4 \Lambda^3} \dfrac{(1+\frac{\alpha}{3})}{\pi^4} \cdot X^2
- \dfrac{f^3}{2 \Lambda^3} \dfrac{X}{\pi^3} \cdot \Box\pi
,
\ee
where $\Lambda$, $f$ and $ \alpha$ are the same parameters
as in the Genesis model in Ref.~\cite{Creminelli2}.
Upon field redefinition
$\phi = f\cdot \log\left(-\sqrt{\frac{3f}{2\Lambda^3}}\frac{1}{\pi}\right)$
the action~\eqref{lagr_asymp_early}
coincides with that in Ref.~\cite{Creminelli2}.
Note that the non-zero
parameter $\alpha$ ensures the subluminal propagation
of scalar modes during the Genesis stage. We confirm
this explicitly
below, see Fig.~\ref{sound_speed}.

On the other hand,
we require that the solution boils down, at late times  $t \to+\infty$, to
a standard flat FLRW Universe driven by a conventional
massless scalar field. This late epoch has
the following Hubble parameter:
\be
\label{Hubble_late}
% &&t \to -\infty: \quad H \simeq ... , \quad a(t) \simeq ...,\\
t \to +\infty: \quad H \simeq \dfrac{1}{3t},
\ee
and the Lagrangian reads
\be
\label{lagr_asymp_late}
\mathcal{L}_{t\to+\infty} = -\frac12 R + \dfrac{X}{3\pi^2},
\ee
which indeed implies that $\phi = \sqrt{\frac23} \log(\pi)$
is a conventional massless scalar field.

Our (admittedly,
fairly arbitrary) choice of the Hubble parameter with
asymptotic behaviour~\eqref{Hubble_early} and~\eqref{Hubble_late} is
\be
\label{Hubble}
H(t) = \left[\left(
4\; \frac{\Lambda^3}{f^3} \cdot  \frac{t^2 \:(1- \mbox{tanh}(t/\tau))}{2\left(1+\alpha/3\right)} + 3 \cdot \frac{1 + \mbox{tanh}(t/\tau)}2
\right)\sqrt{2\tau^2 +t^2}\right]^{-1},
\ee
where $\tau$ is a constant which
controls the characteristic  time scale.
In what follows we take $\tau \gg 1$ to make this
scale
safely greater than Planck time.

In order to reconstruct the Lagrangian of beyond Horndeski theory,
which admits the solution~\eqref{rolling_pi}, \eqref{Hubble},
we utilize the following Ansatz for the Lagrangian functions
in~\eqref{lagrangian}:
\begin{subequations}
\label{Ansatz}
\begin{align}
\label{F_ansatz}
& F(\pi, X) = f_1(\pi)\cdot X + f_2(\pi)\cdot X^2 + f_3(\pi)\cdot X^3, \\
\label{K_ansatz}
& K(\pi, X) = k_1(\pi)\cdot X, \\
\label{G4_ansatz}
& G_4(\pi, X) = \frac12 + g_{40}(\pi) + g_{41}(\pi) \cdot X,\\
%& G_5(\pi, X) = 0, \\
\label{F4_ansatz}
& F_4(\pi, X) = f_{40}(\pi).
%&  F_5(\pi, X) = 0,
\end{align}
\end{subequations}
The central point of the reconstruction procedure is to find the
explicit forms of functions $f_i(\pi)$, $k_1(\pi)$,
$g_{40}(\pi)$ and $f_{40}(\pi)$ by satisfying the stability
conditions~\eqref{stability_constraints} and background Einstein
equations~\eqref{einstein}. At the same time, the
behaviour of these Lagrangian functions as $t\to\pm\infty$ must comply with the
asymptotics~\eqref{lagr_asymp_early} and~\eqref{lagr_asymp_late}.

Let us  describe the algorithm for finding the functions
in \eqref{Ansatz} for a specific solution~\eqref{rolling_pi},
\eqref{Hubble}. We write $\mathcal{D}$, $\mathcal{G_T}$, $\mathcal{F_T}$,
$\Sigma$ and $\Theta$
(see eqs.~\eqref{list_coeff_setup}), which are involved
in the stability criteria~\eqref{stability_constraints},
in terms of $f_i(t)$, $k_1(t)$, etc.:
\begin{subequations}
\begin{align}
\label{D_ansatz}
 \mathcal{D} &= 2f_{40}(t),\\
 %%%%%%
\label{Gt_ansatz}
 \mathcal{G_T} &= 1 + 2 g_{40}(t) - 2 g_{41}(t) - 2 f_{40}(t),\\
 %%%%%%
\label{Ft_ansatz}
 \mathcal{F_T} &= 1 + 2 g_{40}(t) + 2 g_{41}(t),\\
 %%%%%%
\label{Sigma_ansatz}
\Sigma &= f_1(t) + 6 f_2(t) + 15 f_3(t) - 3 H^2 + 168 f_{40}(t) H^2 - 6 g_{40}(t) H^2  \\\nonumber&
 + 36 g_{41}(t) H^2 + 12 H k_1(t) -
 6 H \dot{g}_{40}(t) - 36 H \dot{g}_{41}(t) -
 2 \dot{k}_1(t),\\
%%%%%%
\label{Theta_ansatz}
 \Theta &= H - 14 f_{40}(t) H + 2 g_{40}(t) H - 6 g_{41}(t) H - k_1(t) +
 \dot{g}_{40}(t) + 3 \dot{g}_{41}(t),
\end{align}
\end{subequations}
where $t$ is identified with  $\pi$
in accordance with \eqref{rolling_pi}. 
The Einstein equations~\eqref{einstein} in terms
of the Ansatz functions~\eqref{Ansatz} read
\bea
\label{eq:00einsteinAnsatz}
&-f_1(t)-3 f_2(t)-5 f_3(t)+3 H^2-42 f_{40}(t) H^2+6 g_{40}(t) H^2
-18 g_{41}(t) H^2-6 H k_1(t)
\\\nonumber&
\phantom{\qquad\qquad\qquad\qquad\qquad\qquad\qquad\qquad}
+6 H \dot{g}_{40}(t)+18 H \dot{g}_{41}(t)
+\dot{k}_1(t) = 0,\\\nonumber
\\
\label{eq:ijeinsteinAnsatz}
&f_1(t)+f_2(t)+f_3(t)+3 H^2-6 f_{40}(t) H^2+6 g_{40}(t) H^2-6 g_{41}(t) H^2-4 H \dot{f}_{40}(t)
\\\nonumber&
+4 H \dot{g}_{40}(t)
-4 H \dot{g}_{41}(t)+2 \dot{H} -4 f_{40}(t) \dot{H} 
+4 g_{40}(t) \dot{H}
-4 g_{41}(t) \dot{H}-\dot{k}_1(t)
\\\nonumber&
\qquad\qquad\qquad\qquad\qquad\qquad\qquad\qquad\qquad\qquad\quad
+2 \ddot{g}_{40}(t)+2 \ddot{g}_{41}(t) = 0.
\eea
These expressions will be used in what follows.

First, for the sake of simplicity,
we choose
\be
\label{GtFtchoice}
\forall t: \quad \mathcal{G_T} = \mathcal{F_T} = 1, \quad c_\mathcal{T}^2 = 1,
\ee
which guarantees the absence of ghosts and gradient
instabilities in the tensor sector, as well as strictly luminal
propagation of  gravitational waves. The latter choice appears
natural since both asymptotics~\eqref{lagr_asymp_early}
and~\eqref{lagr_asymp_late} have $G_4(\pi,X) \to 1/2$
(i.e., $g_{40}(t) \to 0$ and $g_{41}(t) \to 0$) and
$F_4(\pi,X) = f_{40}(t) \to 0$ as $t \to \pm\infty$,
which, according to eqs.~\eqref{Gt_ansatz} and~\eqref{Ft_ansatz} gives
$\mathcal{G_T}|_{ t \to \pm\infty} = 1$ and
$\mathcal{F_T}|_{ t \to \pm\infty} = 1$.
Second, we ensure that the solution is free of gradient
instabilities in the scalar sector at all times, i.e.,
the inequality~\eqref{nogo1} holds during the entire evolution.
In order to evade the no-go theorem and allow $\xi$ to cross zero,
we choose
\be
\label{f40}
\dfrac12 \mathcal{D} = f_{40}(t) =  - w\cdot \mbox{sech}^2\left(\frac{t}{\tau}+u\right),
\ee
where parameters $w$ and $u$ are introduced so that
$(\mathcal{G_T}+\mathcal{D}\dot{\pi})$ in~\eqref{nogo1}
crosses zero twice (single zero-crossing or touching
zero corresponds to a fine-tuned case, see Ref.~\cite{bounceI}
for
discussion). The choice made in eq.~\eqref{f40}
completely defines $F_4(\pi,X)$
in~\eqref{F4_ansatz}, which rapidly vanishes as
$t \to \pm\infty$ in full accordance
with the required  asymptotics. By making use
of~\eqref{GtFtchoice} together with eqs.~\eqref{Gt_ansatz}
and~\eqref{Ft_ansatz}, we find $g_{40}(t)$ and
$g_{41}(t)$:
\be
\label{g40_g41}
g_{40}(t) = - g_{41}(t) = - \frac{w}{2} \;\mbox{sech}^2\left(\frac{t}{\tau}+u\right).
\ee
This completes the reconstruction of $G_4(\pi,X)$ in~\eqref{G4_ansatz}.

Let us now take care of $\gamma$-crossing (the property that
$\Theta$ crosses zero).
With the asymptotic forms of the Lagrangian
in eqs.~\eqref{lagr_asymp_early} and~\eqref{lagr_asymp_late},
the asymptotics of $\Theta$ are
as follows (see eq.~\eqref{eq:Theta_coeff_setup}):
\be
\label{Theta_asymp}
\Theta|_{t \to -\infty} \to -\dfrac{1}{|t|^3}, \quad
 \Theta|_{t \to +\infty} \to \dfrac{1}{3t}.
 \ee
 Note the opposite signs in opposite asymptotics,
 as anticipated in Sec.~\ref{sec:stable}.
 A possible choice  for $\Theta$ is then
 \be
\label{Theta_choice}
\Theta = \frac{t}{3(t^2+\tau^2)+\frac{2 t^4 \Lambda^3 (1 - \mbox{\scriptsize{tanh}}(\frac{t}{\tau}))}{f^3 (1 - \frac{\alpha}{3})}}.
\ee
With this choice of $\Theta$ and our form of $\mathcal{D}$ in~\eqref{f40} (and $\mathcal{G_T}=1$),
the function  $\mathcal{F_S}  $ given by \eqref{eq:FS_setup} is positive
at all times.
According to eq.~\eqref{Theta_ansatz}, $\Theta$
is related to a yet undefined function
$k_1(t)$. For our choice of $\Theta$ in
eq.~\eqref{Theta_choice}, $k_1$ reads
\be
\label{eq:k1}
k_1(t) = - \Theta+H +\frac{1}{3}\frac{1}{\cosh^2\left(\frac{t}{\tau}+u\right)}\left[30wH-6\frac{w}{\tau}\mbox{tanh} \left(\frac{t}{\tau}+u\right)\right] \; .
\ee
This completely determines $K(\pi,X)$ through~\eqref{K_ansatz}.

Finally, still undetermined
functions $f_1(t)$, $f_2(t)$, $f_3(t)$
in~\eqref{F_ansatz} are chosen in such a way that  the
background Einstein equations~\eqref{eq:00einsteinAnsatz} and~\eqref{eq:ijeinsteinAnsatz} are satisfied, and the
remaining stability condition
$\mathcal{G_S} \geq \mathcal{F_S}$ holds (recall that $ \mathcal{F_S} > 0$
by the above construction).
Einstein equations~\eqref{eq:00einsteinAnsatz} and~\eqref{eq:ijeinsteinAnsatz}
enable us
to express $f_1(t)$ and $f_2(t)$ in terms of
already defined functions $g_{40}$, $g_{41}$, $f_{40}$, $k_1$
and the unknown $f_3(t)$ as follows:
\bea
\label{eq:f1}
 f_1(t) &= f_3(t) - 6 H^2 + 30 f_{40}(t) H^2 - 12 g_{40}(t) H^2 +
 18 g_{41}(t) H^2 
 \nonumber\\& + 3 H k_1(t) + 6 H \dot{f}_{40}(t) -
 9 H \dot{g}_{40}(t) - 3 H \dot{g}_{41}(t) -
 3 \dot{H} 
 \nonumber\\& + 6 f_{40}(t) \dot{H} -
 6 g_{40}(t) \dot{H} + 6 g_{41}(t) \dot{H} +
 \dot{k}_1(t) - 3 \ddot{g}_{40}(t) -
 3 \ddot{g}_{41}(t), \\
 \nonumber\\
 \label{eq:f2}
 f_2(t) &=-2 f_3(t) + 3 H^2 - 24 f_{40}(t) H^2 + 6 g_{40}(t) H^2 -
 12 g_{41}(t) H^2 \nonumber\\& - 3 H k_1(t) - 2 H \dot{f}_{40}(t) +
 5 H \dot{g}_{40}(t) + 7 H \dot{g}_{41}(t) +
 \dot{H} \nonumber\\& - 2 f_{40}(t) \dot{H} +
 2 g_{40}(t) \dot{H} -
 2 g_{41}(t) \dot{H} + \ddot{g}_{40}(t) +
  \ddot{g}_{41}(t).
\eea
The only free function left is $f_3(t)$, which is utilized to
make sure that the solution is not only free of ghosts in the
scalar sector, but also that the scalar modes are safely subluminal.
This is done by adjusting the behaviour of $\Sigma$ in
eq.~\eqref{eq:GS_setup}, which, according to eq.~\eqref{Sigma_ansatz},
involves the leftover $f_3(t)$. We take $\Sigma$
in the following form:
\be
\label{Sigma_choice}
\Sigma = \dfrac{3f^3}{4\Lambda^3} \dfrac{1+\alpha}{(\tau^2 +t^2)^2},
\ee
which agrees with the asymptotics  required
by~\eqref{lagr_asymp_early} as $t \to -\infty$ and, at the same time,
is sufficient to suppress the first term in eq.~\eqref{eq:GS_setup}
as $t \to +\infty$, leading to $\mathcal{G_S} \to 3 \mathcal{G_T}$ .
Together with the previously determined $\mathcal{F_S}$ in
eqs.~\eqref{GtFtchoice},~\eqref{f40} and~\eqref{Theta_choice},
the
behaviour of $\mathcal{G_S}$  is sufficient
to have at most luminal propagation of the scalar modes,
$c_{\mathcal{S}}^2 \leq 1$. Hence, by specifying $\Sigma$
  in eq.~\eqref{Sigma_choice} and using
  eqs.~\eqref{Sigma_ansatz}, \eqref{eq:f1} and
  \eqref{eq:f2} we obtain $f_3(t)$ in the following form:
\be
f_3(t) = \frac14 \left(\Sigma + 3 H k_1(t)+9 H^2 
\left[\frac{8 w}{\cosh^2\left(\frac{t}{\tau}+u\right)}-1\right]
-3 \dot{H}+\dot{k}_1(t)\right),
\ee
where $k_1(t)$ can be read off in eq.~\eqref{eq:k1}.
This completes the reconstruction
of $F(\pi,X)$ in Ansatz~\eqref{Ansatz}.

\begin{figure}[h!]\begin{center}\hspace{-1cm}
{\includegraphics[width=0.5\linewidth]{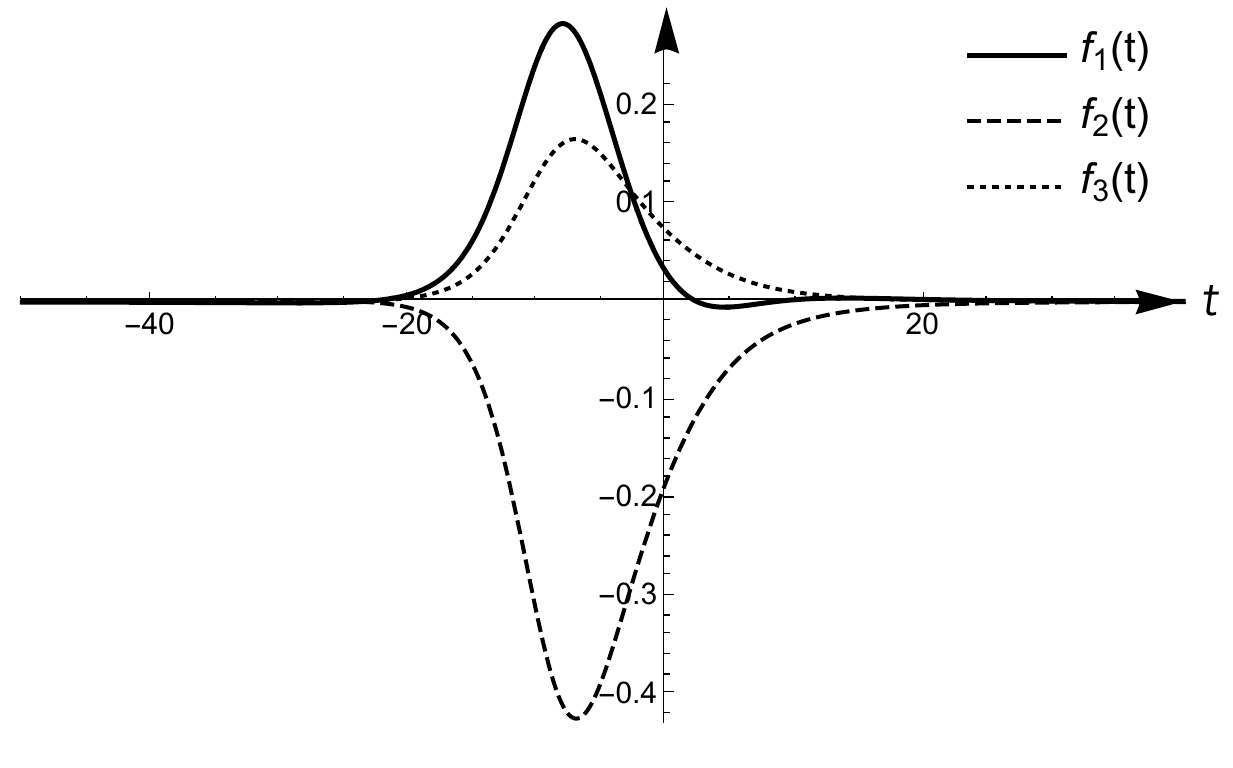}}\hspace{2.8cm}\hspace{-3cm}
{\includegraphics[width=0.5\linewidth] {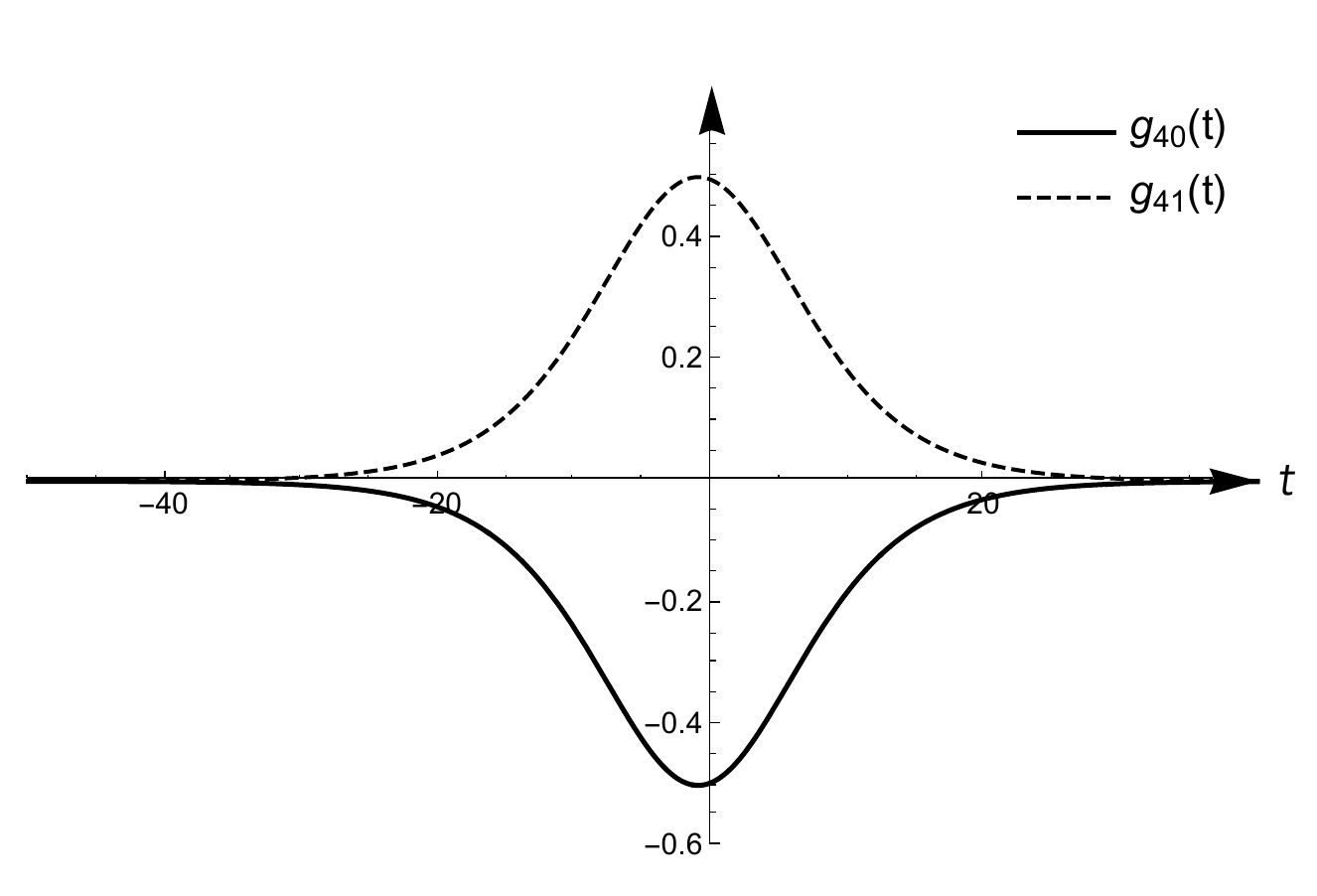}}

\vspace{-2cm}
{\includegraphics[width=0.5\linewidth]
{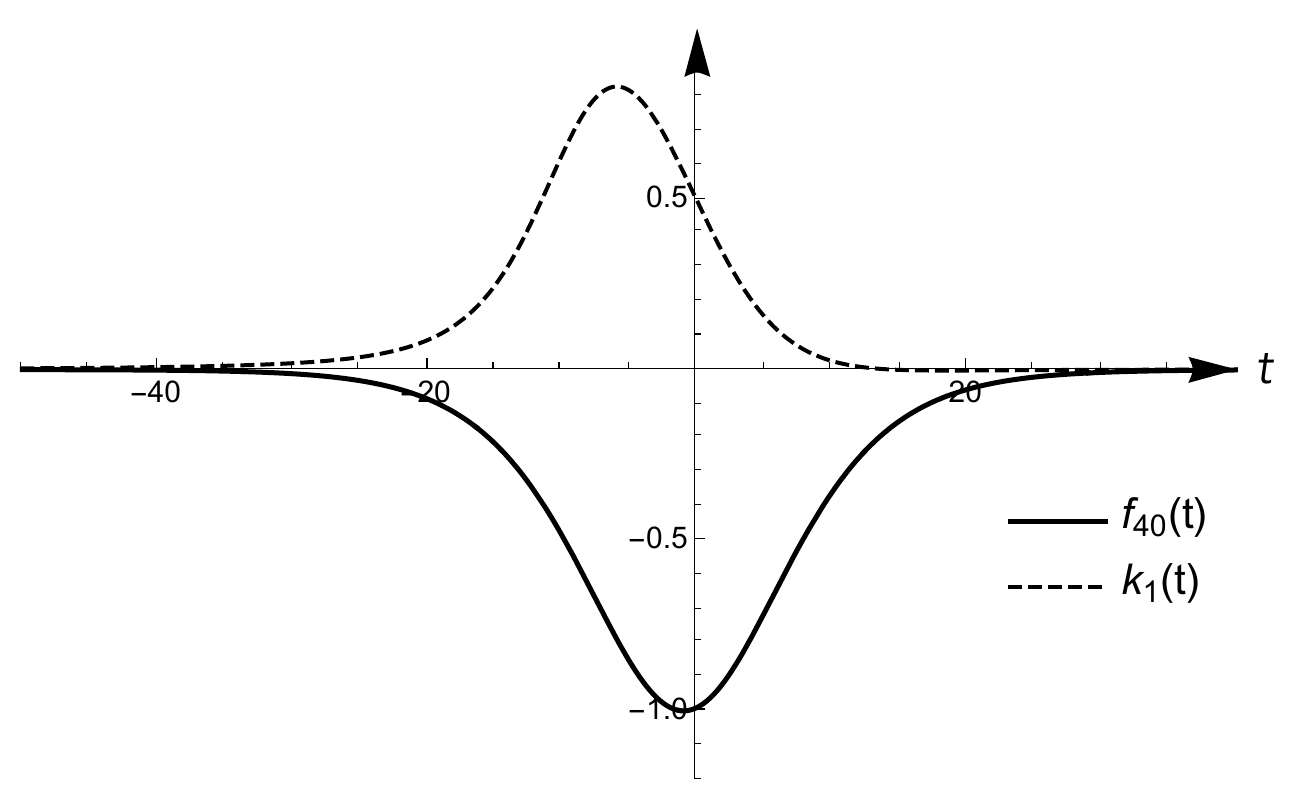}}\hspace{1cm}
\caption{\footnotesize{The Lagrangian functions $f_0(t)$,
$f_1(t)$, $f_2(t)$, $g_{40}(t)$, $g_{41}(t)$,
$f_{40}(t)$ and $f_{41}(t)$, with the following choice of
parameters:
$u=1/10$, $w=1$, $\tau = 10$, $\Lambda = \alpha =1$ and $f=10$. This choice guarantees that
the solution is not fine-tuned and its characteristic time
scale safely exceeds the Planck time.
}} \label{LagrangianFunctions}
\end{center}\end{figure}

\begin{figure}[h]
\center{\hspace{-0.1cm}\includegraphics[width=0.64\linewidth]{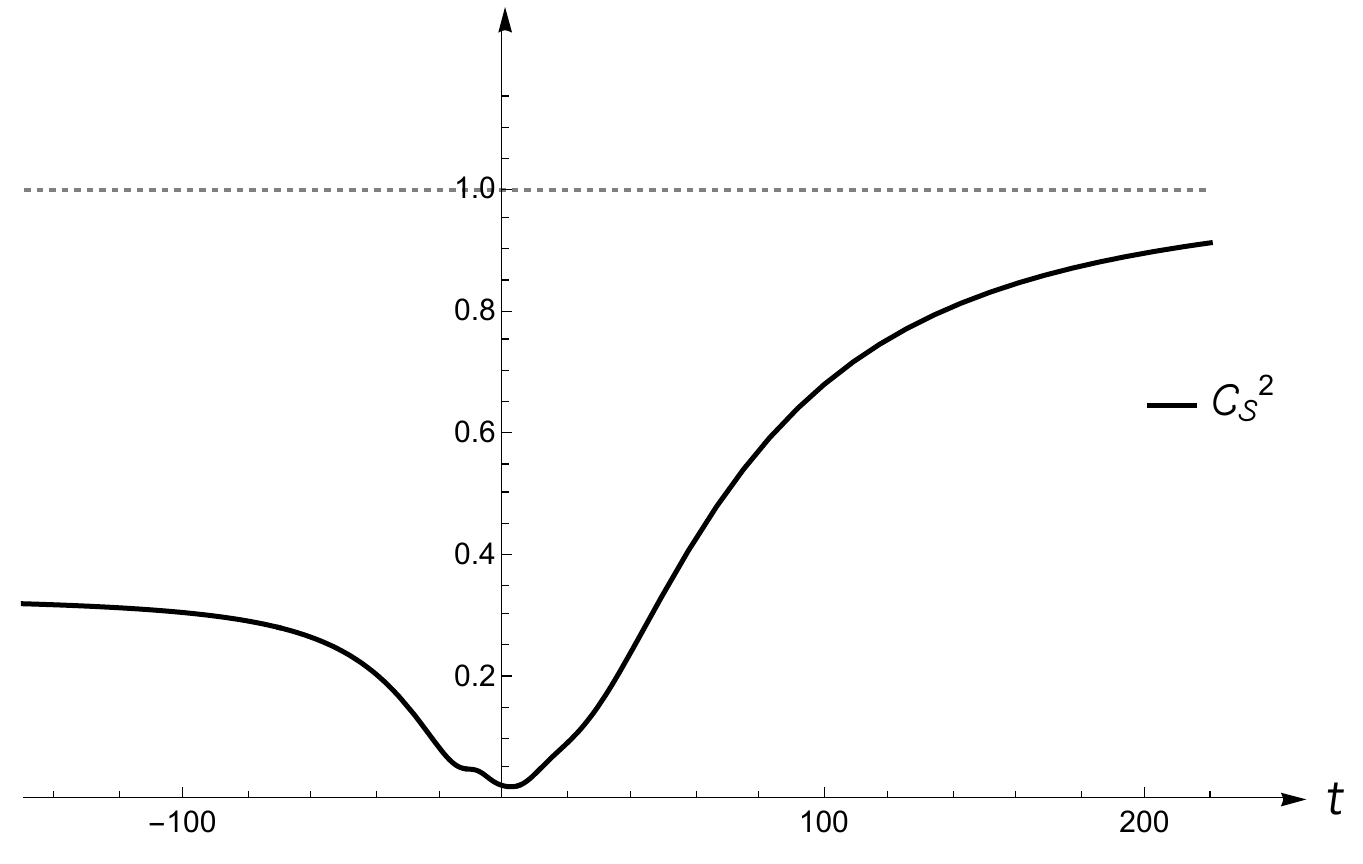}
% \\
% \hspace{1cm}
% \includegraphics[width=0.4\linewidth]{xitheta}
}
\caption{The sound speed squared of the scalar perturbations is
non-negative for all times and asymptotically tends to 1 in
the infinite future.
The parameters $u$, $w$, $\tau$, $\Lambda$, $\alpha$ and $f$ are the same as in Fig.~\ref{LagrangianFunctions}.}
\label{sound_speed}
\end{figure}

The reconstructed functions $f_1(t)$, $f_2(t)$, $f_3(t)$,
$k_1(t)$, $g_{40}(t)$, $g_{41}(t)$ and $f_{40}(t)$ are  shown
in Fig.~\ref{LagrangianFunctions}.
%%%
%%%
%%%
%%%
Their asymptotic behaviour as $t \to -\infty$ is as follows:
\be
\label{asymptoticspast}
f_1(t) = f_2(t) \propto \frac{1}{t^4}, \;\;\;\;
f_3(t) \propto \frac{1}{ t^6},  \;\;\;\;
k_1(t) \propto \frac{1}{ t^3},  \;\;\;\;
g_{40}(t) = g_{41}(t) = f_{40}(t) \propto e^{2t/\tau}.
\ee
As promised, the beyond Horndeski function
$F_4(\pi,X)$ decreases significantly faster as $t \to -\infty$ as
compared to $F(\pi,X)$ and $K(\pi,X)$, while $F(\pi,X)$ and
$K(\pi,X)$ have the power-law behaviour dictated by
\eqref{lagr_asymp_early}.
The functions $g_{40}(t)$ and $g_{41}(t)$ vanish exponentially,
which corresponds to GR during the Genesis stage, in full
accordance with the asymptotic~\eqref{lagr_asymp_early}.

As $t \to +\infty$, we have
\be
\label{asymptoticspast}
f_1(t) = \frac{1}{ 3 t^2},  \;\;\;\;
f_2(t) = f_3(t) \propto \frac{1}{t^4}, \;\;\;\;
k_1(t) \propto \frac{1}{ t^5},  \;\;\;\;
g_{40}(t) = g_{41}(t) = f_{40}(t) \propto e^{-2t/\tau},
\ee
which corresponds to the required form of the Lagrangian at
late times given by eq.~\eqref{lagr_asymp_late}.

\begin{figure}[h]
\center{\includegraphics[width=0.9\linewidth]{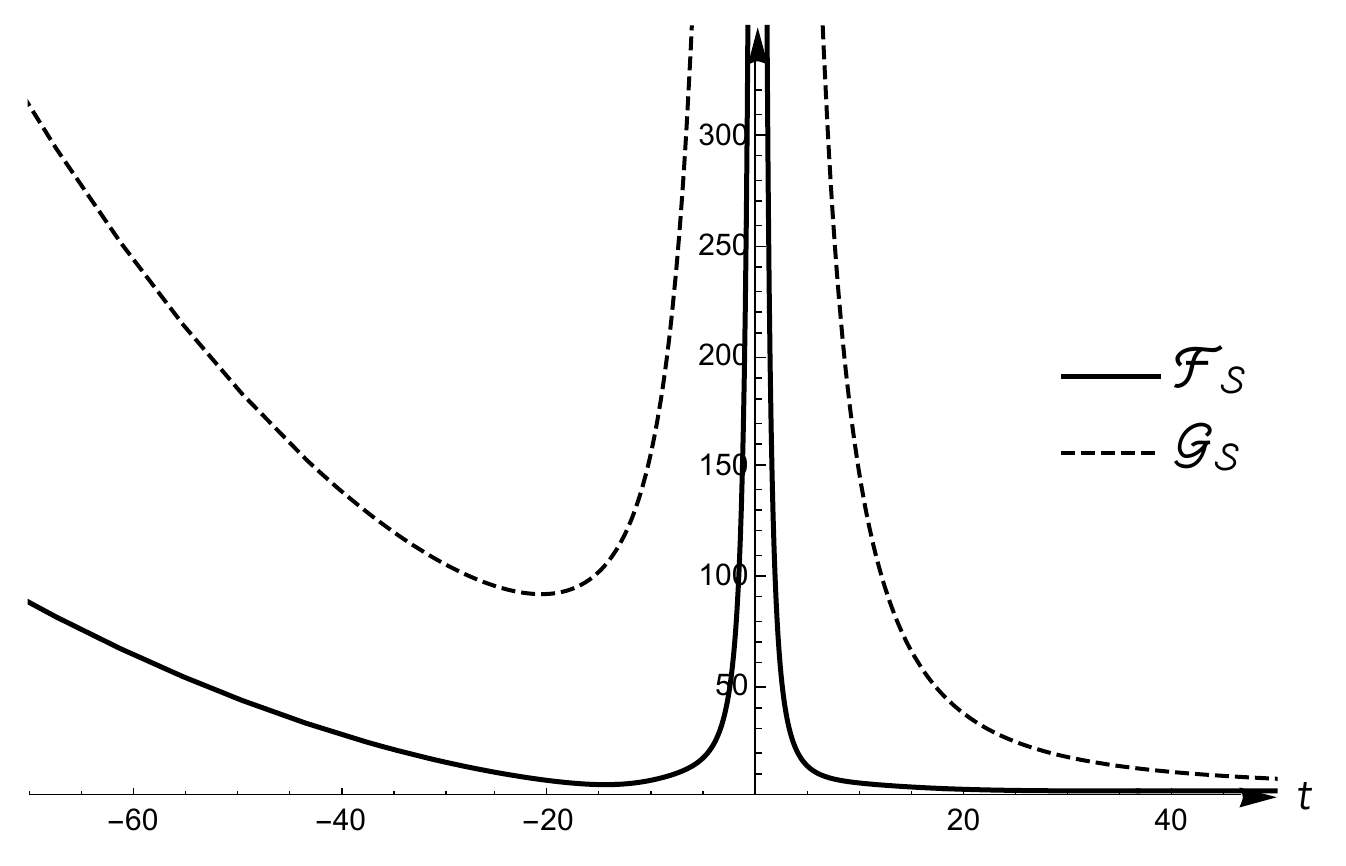} }
\caption{The coefficients $\mathcal{G_S}$ and $\mathcal{F_S}$;
the parameters $u$, $w$, $\tau$, $\Lambda$, $\alpha$ and $f$ are the same as in
Fig.~\ref{LagrangianFunctions}.}
\label{plot_gsfs}
\end{figure}

We show the coefficients $\mathcal{G_S}$ and $\mathcal{F_S}$
responsible for the stability of the scalar sector in Fig.~\ref{plot_gsfs}.
%%%
%%%
%%%
%%%
%%%
%%%
%%%
%%%
\begin{figure}[h!]
\center{\includegraphics[width=0.64\linewidth]{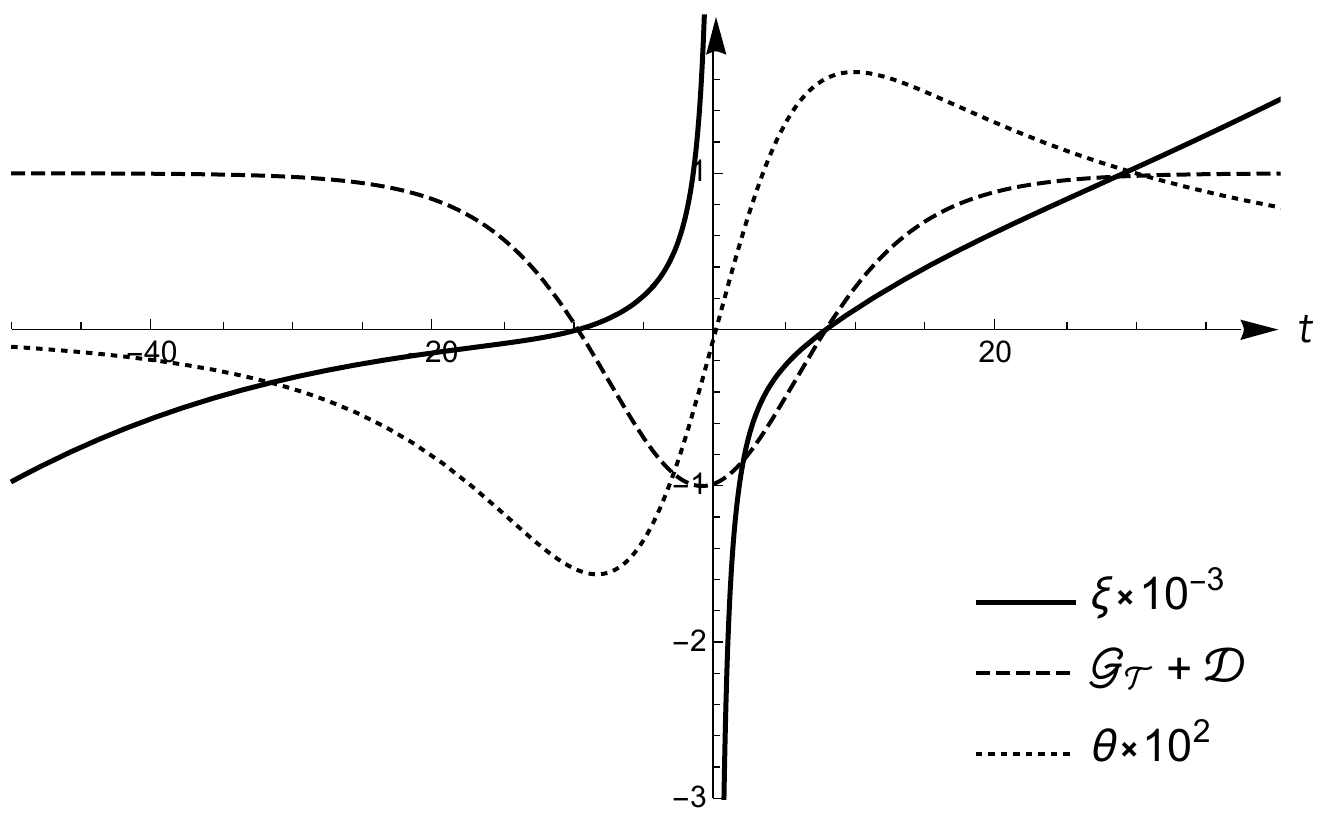}
}
\caption{The plots of $\Theta$, $(\mathcal{G_T}+\mathcal{D})$
and $\xi$ in our model; these functions
play the crucial role in evading the no-go theorem.}
\label{nogo}
\end{figure}
%%%
%%%
The scalar sound speed squared is given in Fig.~\ref{sound_speed};
it confirms the subluminal propagation of perturbations
at early times and reveals that
$c_{\mathcal{S}}^2|_{t\to+\infty} \to 1$,
as expected for
the massless scalar field, at late times. Let us recall that we have
chosen $\mathcal{G_T} = \mathcal{F_T} = 1 $, and hence
$c_{\mathcal{T}}^2 = 1$.

We plot the functions
$\xi$, $(\mathcal{G_T}+\mathcal{D})$
and $\Theta$ in Fig.~\ref{nogo} to clarify the way we evade
the no-go theorem with
our solution and ensure that the
inequality~\eqref{nogo1} holds.

Hence, the reconstructed beyond Horndeski Lagrangian is an explicit example of the theory admitting a complete, stable Genesis solution with both asymptotics described by GR. The solution is indeed free of
instabilities of all kinds and does not suffer from superluminal modes.

%%%%%%%%%%%%%%%%%%%%%%%%%%%%%%%%%%%%%%%%%%%%%%%%%%%%%%%%%%%%%%%%%%%%%%%%%
\section{Conclusion}
\label{sec:conclusion}
In this work, we have revisited the Genesis scenario in beyond
Horndeski theory and suggested a modified version of it.
We have constructed
a specific Lagrangian of beyond Horndeski type,
which admits the completely stable solution with the Genesis epoch
at early times and both asymptotics described by GR as $t \to\pm\infty$.
Unlike the previous version of the scenario suggested
in Ref.~\cite{bounce}, the dynamics during the Genesis stage
is similar to that in the original Genesis model
of Ref.~\cite{Creminelli2}
and is driven by the cubic Galileon, while at late times the theory
tends to
GR + a conventional massless scalar field.
The novel feature is
the
simple behaviour of the theory in both
the asymptotic
past and future, which
results from
allowing $\gamma$-crossing in our model.
We have strengthened the point raised in Refs.~\cite{bounce,bounceI}
that $\gamma$-crossing is the key to constructing  ever-stable
non-singular solutions
with both asymptotics described by GR.
The
stability of the Genesis solution as well as the
required form of asymptotics are explicitly established and
follow from the
reconstruction procedure.
Our judicial choice of the Lagrangian also ensured safe subluminal
or at most luminal propagation of both scalar and tensor modes
at all times.
It is worth noting that in our model, tensor modes propagate 
at the speed of light, which is safe from the observational
viewpoint.
Moreover, since long enough after the Genesis epoch the theory reduces to
that of a conventional massless scalar 
field and GR, the late-time cosmological behavior
is the standard hot stage (provided, of course, that the energy density of
our scalar is converted into heat), so no constraints on our
Lagrangian functions emerge.
The suggested Genesis solution with the ascribed set of properties
is a promising candidate for describing the early time evolution
within the realistic cosmological
models.

%%%%%%%%%%%%%%%%%%%%%%%%%%%%%%%%%%%%%%%%%%%%%%%%%%%%%%%%%%%%%%%%%%%%%%%%%
\section{Acknowledgements}
The work has been supported by Russian Science Foundation Grant No. 19-12-00393.

%%%%%%%%%%%%%%%%%%%%%%%%%%%%%%%%%%%%%%%%%%%%%%%%%%%%%%%%%%%%%%%%%%%%%%%%%

\end{document}